\documentclass[12pt,twoside,reqno]{amsart}
\usepackage{times}
\usepackage{mathptmx}
\usepackage{amssymb}
\begin{document}
\setcounter{page}{1}

\vspace*{1.0cm}

\title[Investigation of Power-Law Damping/Dissipative Forces]{Investigation of Power-Law Damping/Dissipative Forces}
\author{  Ronald E. Mickens$^1$, and Kale Oyedeji$^{2,*}$   }
\date{}
\maketitle

\vspace*{-0.2cm}

\begin{center}
{\footnotesize  $^1$Department of Physics, Clark Atlanta University, Atlanta, GA 30314, USA\\
$^2$Department of Physics, Morehouse College, Atlanta, GA 30314-3773, USA\\

 }

\end{center}
\vskip 2mm

{\footnotesize \noindent {\bf Abstract.} The properties of a one space-dimension, one particle dynamical system under the influence of a purely dissipative force are investigated. Assuming this force depends only on the velocity, it is demonstrated, in contrast to the case of linear damping, that there exist dissipative forces for which the particle \textquotedblleft stops" in a finite time. It is also shown, by an explicit example, that other dissipative forces exist such that they produce dynamics in which the particle achieves zero velocity only after an infinite distance has been traveled. Possible applications of these results to more
complex situations are discussed.}

\vskip 6mm







\renewcommand{\thefootnote}{}
\footnotetext{ $^*$'Kale Oyedeji
\par
E-mail addresses:rmickens@cau.edu,and kale.oyedeji@morehouse.edu
\par
}

\section {Introduction}
\vskip 6mm

 Dissipation and/or damping forces play very significant roles in the
modeling and analysis of many physical systems. \cite{1,2,3,4,5,6}
Dissipation is generally the term used for situations where a conservative
system is acted upon by friction and/or viscous forces, while damping is
often used in cases where the relevant system undergoes oscillatory motions.
However, in many instances, no such clear distinctions can be made. The
major characterizations of such forces separate them into three broad
classes \cite{2,3,6}: viscous damping, aero-dynamic drag, and dry friction,
although, other possible types of forces may be defined when such concepts
are of value, an example being hysteretic damping. \cite{2} Within the
context of the issues discussed in this presentation, we will simply use the
term \textquotedblleft damping force."

If we consider a localized conservative physical system interacting with an
energy absorbing environment, then the system will evolve in time to a
locally stable stationary state. \cite{7} An example of such a system is an
unforced pendulum, set in motion, in air. Eventually, it reaches a state of
equilibrium where it hangs vertically, with no angular velocity. \cite{2}
However, for this and many other systems, it is standard practice, in first
approximation, to represent the damping force by a term linear in the
velocity. An important, nontrivial, and also nonlinear case is the unforced
Duffing equation. \cite{2,3,4,6} 
\begin{equation*}
m\ddot{x}+2\epsilon \dot{x}+\omega ^{2}x+kx^{3}=0,
\end{equation*}%
where $(m,\epsilon ,\omega ^{2},k)$ are constant parameters, all of which
are taken to be positive, except for $\omega ^{2}$and $k$, which may be of
either sign. \cite{6} (The "dot" is used to indicate the time-derivative,
i.e., $\dot{x}$=$\frac{dx(t)}{dt}$.) The general qualitative features of
solutions to this equation are well known. \cite{2,4,6} For $x(0)\neq 0$
and either $\dot{x}(0)=0$ or $\dot{x}(0)\neq 0$, the Duffing oscillator
undergoes damped oscillations, with the rest or equilibrium state being
achieved only in the limit $t\rightarrow \infty $. However, observations of
actual nonlinear oscillatory systems lead to the conclusion that the
oscillations generally take place only over a finite time interval, $T$. For 
$t>T$, the system is at rest in its equilibrium state, 
\begin{equation*}
x(t)=0,\quad \dot{x}(t)=0,\quad t>T,
\end{equation*}%
and the oscillations have stopped. Other common mathematical forms for the
damping force, which we assume is only a function of the velocity, $\dot{x}$%
, and consists of sums of terms proportional to integer power of $\dot{x}$,
give the same result \cite{2,3,4,6} as for linear damping.

The main goal of this presentation is to determine if damping forces exist
such that, at least in certain model systems, the motion or dynamics ends
after a finite time interval. We show that for a certain class of damping
forces, an affirmative answer can be given. To do this, we study the
dynamics of a one-dimensional system: a particle, acted on by a single,
purely damping force, where this force is taken to be a power of the
velocity, and the value of the power is a non-negative real number. While
the focus is on a particular one-dimensional system, the derived results and
related conclusions are of general validity for more complex systems.

In the next section, we briefly state the various assumptions that must be
placed on a valid damping force and explain how they arise from knowledge
and understanding of actual physical systems. Section III gives the model
differential equation and its solutions for various values of a parameter $%
\alpha $, which characterizes the damping force. The results obtained in
section III and their implications are discussed in section IV. Finally, in
section V, our major conclusions are briefly restated and possible
applications are presented.

It should be noted that the results of this article has direct applicability
and relevance to various topics arising in introductory and advance physics
courses, and in many practical applications. Generally, the effects of
damping are studied by assuming the corresponding force is a linear function
of the velocity. While there are circumstances where such a form may be
justified, \cite{1,2,3,4,5,6} the main reason for using this particular
functional form is that it often leads to linear differential equations
which can then be easily solved by techniques \cite{7} well known to the
students and their teachers, and researchers.

\section{Properties of Damping Forces}

Consider a particle of mass, $m$, moving in the direction of the positive $x$%
-axis, such that at time $t=0$, it is located at $x=0$, with velocity $%
v_{0}>0$. Assume that the only force acting on it is a (viscous) damping
force which depends only on the velocity. The equation of motion, \cite{7}
for $m=1$, is 
\begin{equation}
\ddot{x}=-f(\dot{x}),\quad x(0)=0,\quad \dot{x}(0)=v_{0}>0,  \label{eq2.1}
\end{equation}%
where $f(v)$, $v=\dot{x}$, is the negative of the damping force. Physical
restrictions place the following constraints on $f(v)$:

\begin{itemize}
\item[i)] $f(v)$ is continuous in $v$;

\item[ii)] $f(-v)=-f(v)$;

\item[iii)] $f(v)>0$, if $v>0$.
\end{itemize}

Condition i) is a reasonable mathematical requirement and combined with ii)
gives the result $f(0)=0$. Condition ii) is consistent with the fact that
damping forces change signs if the direction of motion is changed to its
opposite. Finally, iii) is what is expected of any purely damping force,
i.e., it should always act in the direction opposite to the motion for any
velocity.

Another characterization of $f(v)$ is to consider its impact on the energy
of the particle. Without the damping force, the particle (with $m=1)$ has a
total energy equal to its kinetic energy $KE=v^{2}/2$. If Eq.~\eqref{eq2.1}
is rewritten to the system form 
\begin{equation*}
\dot{x}=v,\quad \dot{v}=-f(v),
\end{equation*}%
then 
\begin{equation*}
\frac{d(KE)}{dt}=-vf(v)\leq 0.
\end{equation*}%
Therefore, it follows that the $KE$ decreases to zero, as $t\rightarrow
\infty $, and, as a consequence, the effect of $f(v)$ is to have the energy
continuously decrease to zero.

The class of damping functions investigated here are defined by a parameter $%
\alpha $, and have a power-law dependence on $v$. The form selected is 
\begin{equation}
f(v)=[\lambda \mathop{\rm sgn}(v)]|v|^{\alpha },\quad \lambda >0,\quad
\alpha >0,  \label{eq2.2}
\end{equation}%
where $\lambda $ is a positive constant, $\alpha $ is real and positive, and 
$\mathop{\rm sgn}(v)$ is the \textquotedblleft sign-function" \cite{8}, 
\begin{equation*}
\mathop{\rm sgn}(v)=%
\begin{cases}
+1, & \text{$v>0$}, \\ 
0, & \text{$v=0$}, \\ 
-1, & \text{$v<0$}.%
\end{cases}%
\end{equation*}%
Inspection of Eq.~\eqref{eq2.2} shows that it satisfies the three conditions
i), ii), and iii). However, for our \textquotedblleft toy" physical system,
the $\mathop{\rm
sgn}(v)$ does not need to be indicated since the velocity is positive and
never changes direction, i.e., combining Eqs.~\eqref{eq2.1} and \eqref{eq2.2}%
, we obtain 
\begin{equation}
\ddot{x}=-\lambda (\dot{x})^{\alpha },\quad x(0)=0,\quad \dot{x}(0)=v_{0}>0.
\label{eq2.3}
\end{equation}

\section{Calculations}

The dynamics of Eq.~\eqref{eq2.3} will now be determined. To do so, five
intervals of $\alpha $-values are examined. They are A)~$\alpha >2$, B)~$%
\alpha =2$, C)~$1<\alpha <2$, D)~$\alpha =1$, and E)~$0<\alpha <1$.

\subsection{$\protect\alpha >2$}

Let $\alpha =2+\beta $, where $\beta >0$. Therefore, Eq.~\eqref{eq2.3}
becomes 
\begin{equation}
\dot{v}=-\lambda v^{2+\beta },\quad v(0)=v_{0}>0.  \label{eq3.1}
\end{equation}%
An elementary integration gives the solution \cite{7,8} 
\begin{equation}
v(t)=\frac{v_{0}}{\left[ 1+\lambda (1+\beta )v_{0}^{(1+\beta )}t\right] ^{%
\frac{1}{1+\beta }}}\,.  \label{eq3.2}
\end{equation}%
Note that for large $t$, $v(t)$ is asymptotic to 
\begin{equation}
v(t)\sim \frac{c_{1}}{t^{\left( \frac{1}{1+\beta }\right) }},  \label{eq3.3}
\end{equation}%
where $c_{1}$ is expressible in terms of $\lambda $, $\beta $, and $v_{0}$.
This result implies that 
\begin{equation}
\lim_{t\rightarrow \infty }v(t)=0.  \label{eq3.4}
\end{equation}

The displacement function, $x(t)$, is given by 
\begin{equation}  \label{eq3.5}
x(t)=\int^t_0 v(z)dz,
\end{equation}
which automatically incorporates the condition $x(0)=0$. However, while the
substitution of Eq.~\eqref{eq3.2} into this integral will allow the explicit
calculation of $x(t)$, our interest is in $x(\infty)$, i.e., 
\begin{equation}  \label{eq3.6}
x(\infty)=\lim_{t\to\infty} x(t)\equiv \int^\infty_0 v(z)dz.
\end{equation}
From the calculus \cite{9}, it follows that the result given in Eq.~%
\eqref{eq3.3} implies that the singular integral in Eq.~\eqref{eq3.6} is
unbounded. Thus, while $v(t)\to 0$, as $t\to\infty$, the particle's motion
carries it an infinite distance from the origin.

\subsection{$\protect\alpha=2$}

For this case, the equation of motion is 
\begin{equation}  \label{eq3.7}
\dot v=-\lambda v^2,\quad v(0)=v_0>0,
\end{equation}
and its solution is 
\begin{equation}  \label{eq3.8}
v(t)=\frac{v_0}{1+\lambda v_0t}.
\end{equation}
Likewise, $x(t)$ is 
\begin{equation}  \label{eq3.9}
x(t)=\int^t_0 v(z)dz=\frac{\ln (1+\lambda v_0t)}\lambda\,,
\end{equation}
and $x(\infty)=\infty$.

\subsection{$1<\protect\alpha<2$}

Let $\alpha=1+\gamma$, where $0<\gamma <1$, and the equation of motion is 
\begin{equation}  \label{eq3.10}
\dot v=-\lambda v^{1+\gamma},\quad v(0)=v_0>0;
\end{equation}
an elementary integration gives 
\begin{equation}  \label{eq3.11}
v(t)=\frac{v_0}{(1+\lambda\gamma v^\gamma_0 t)^{1/\gamma}}\,.
\end{equation}
Therefore, 
\begin{equation}  \label{eq3.12}
x(t)=v_0\int^t_0 \frac{dz}{(1+\lambda\gamma v^\gamma_0 z)^{1/\gamma}} =\left[%
\frac{v^{(1-\gamma)}_0}{\lambda(1-\gamma)}\right] \left\{1-\frac1{(1+\lambda%
\gamma v^\gamma_0 t)^{1-\gamma/\gamma}}\right\},
\end{equation}
and 
\begin{equation}  \label{eq3.13}
x(\infty)=\frac{v^{(1-\gamma)}_0}{\lambda(1-\gamma)}<\infty.
\end{equation}
Thus, it follows that the velocity decreases to zero and the distance
traveled by the particle is finite.

\subsection{$\protect\alpha =1$}

For this case, 
\begin{equation}  \label{eq3.14}
\dot v=-\lambda v,\quad v(0)=v_0>0,
\end{equation}
and the solutions for $v(t)$ and $x(t)$ are, respectively, 
\begin{equation}  \label{eq3.15}
v(t)=v_0 e^{-\lambda t},
\end{equation}
\begin{equation}  \label{eq3.16}
x(t)=\left(\frac{v_0}\lambda\right)(1-e^{-\lambda t}).
\end{equation}
The velocity and displacement, respectively, monotonically decrease and
increase to the values $v(\infty)=0$ and $x(\infty)=v_0/\lambda$.

\subsection{$0<\protect\alpha <1$}

Let $\alpha =1-\delta $, where $0<\delta <1$. Then the equation of motion is 
\begin{equation}
\dot{v}=-\lambda v^{1-\delta },\quad v(0)=v_{0}>0  \label{eq3.17}
\end{equation}%
and its formal solution is 
\begin{equation*}
v(t)=[v_{0}^{\delta }-\delta \lambda t]^{1/\delta }.
\end{equation*}%
Note that this $v(t)$ can become complex valued for $t>t^{\ast
}=v_{0}^{\delta }/\delta \lambda $ and also observe that since Eq.~%
\eqref{eq3.17} is a nonlinear differential equation, $v(t)=0$, is also a
valid, nontrivial solution. \cite{7} The implication of these facts is that $%
v(t)$ is expressed as the following piecewise, continuous function 
\begin{equation}
v(t)=%
\begin{cases}
\lbrack v_{0}^{\delta }-\lambda \delta t]^{1/\delta }, & \text{$0<t\leq
t^{\ast }=\dfrac{v_{0}^{\delta }}{\delta \lambda }$}, \\ 
0, & \text{$t>t^{\ast }$}.%
\end{cases}
\label{eq3.18}
\end{equation}%
Likewise, $x(t)$ is 
\begin{subequations}
\label{eq3.19}
\begin{align}
x(t)&=\left[\frac1{\lambda (1+\delta)}\right] \left\{ v^{1+\delta}_0 -\left[%
v^\delta_0-\lambda\delta t \right]^{\frac{1+\delta}\delta} \right\},
\label{eq3.19a} \\
\intertext{\centerline{$0<t\le t^*$}}
\intertext{and}
x(t)&= \frac{v^{1+\delta}_0}{\lambda (1+\delta)},\quad t>t^*.
\label{eq3.19b}
\end{align}
Consequently, the particle stops moving in a finite time, after having
traveled the finite distance given in Eq.~\eqref{eq3.19b}

\section{Results and Discussion}

We now summarize the results obtained in section III. Several of the
outcomes are surprising and have not been either explicitly stated and/or
discussed in any of the literature known to the authors.

To begin, observe that both of the following differential equations are
mathematically equivalent: 
\end{subequations}
\begin{subequations}
\label{eq4.1}
\begin{alignat}{2}
\ddot{x}& =-[\lambda \mathop{\rm sgn}(\dot{x})](\dot{x})^{\alpha }, & &
\quad x(0)=0,\quad \dot{x}(0)=v_{0}>0,  \label{eq4.1a} \\
\dot{v}& =-[\lambda \mathop{\rm sgn}v](v)^{\alpha }, & & \quad v(0)=\dot{x}%
(0)=v_{0}>0,  \label{eq4.1b}
\end{alignat}%
since $v=\dot{x}$. In the discussion below, both forms will be used. Also,
it should be noted that the cases where $\alpha =1$ or 2 are used quite
extensively in the modeling of physical systems where damping plays an
important role. The $\alpha =1$ value is applied to situations where
so-called \textquotedblleft linear viscous damping" occurs and is used often
in the modeling of nonlinear vibrations and oscillations. \cite{1,2,3,4,6}
However, $\alpha =2$, i.e., quadratic damping, generally appears in fluid
dynamic problems at high Reynold's number; \ \cite{2,3,5,6} also see Ref.~%
\cite{6}, (section 6.2.2), for more details and additional references
to the appropriate literature.

The $\alpha =0$ case provides a model of dry friction; in general, this type
of friction occurs when two solid surfaces are in contact and moving
relative to each other. \cite{2,3,4} This case is not considered in our
studies since the full, anti-symmetric damping force corresponding to this
situation is 
\end{subequations}
\begin{equation*}
f(v)=\lambda \mathop{\rm sgn}(v),\quad \alpha =0,
\end{equation*}%
and consequently is not continuous at $v=0$. However, by considering a
function such as 
\begin{equation*}
\dot{v}=-\lambda \tanh (av),
\end{equation*}%
where $a$ is a positive parameter, and then taking the limit as $%
a\rightarrow \infty $, the $\alpha =0$ case is reproduced.

We now summarize the results of the calculations presented in section III;
for all cases, the initial conditions are those indicated in Eqs.~%
\eqref{eq4.1}:

a) For $\alpha \ge 2$, the velocity, $v(t)$, monotonically decreases from $%
v(0) =v_0>0$ to zero, while $x(t)$ monotonically increases to $%
x(\infty)=\infty$, where $x(\infty)=\lim (t\to\infty)x(t)$.

b) If $1\leq \alpha <2$, then $v(t)$ monotonically decreases from $%
v(0)=v_{0}>0$ to zero. However, $x(t)$ achieves a finite maximum value,
i.e., 
\begin{equation*}
\lim_{t\rightarrow \infty }x(t)=x_{\max }<\infty ,
\end{equation*}%
and it takes an infinite length of time to reach this limiting value.

c) The $\alpha$ values in the interval, $0<\alpha <1$, all give rise to
motions for which $v(t)$ decreases and $x(t)$ increases; however, both
variables reach their limiting values in a finite time, i.e., there exists a
time, $t^*$, such that $v(t^*)=0$ and $v(t)$, for $t>t^*$, is zero; and $%
x(t) $ increases to a maximum value $x_{\max}$ at $t=t^*$ and for $t>t^*$, $%
x(t)=x_{\max}$.

The third case is the interesting and important one. Since actual
conservative physical systems, acted upon by only external damping forces,
are observed/found to reach a stable equilibrium state in a finite time,
then if our results are of general applicability, it follows that for the
class of damping forces considered here, only the $\alpha $-parameter
interval, $0<\alpha <1$, should be used to model the required dissipative or
damping force for such systems. Further, because most dynamic models are
represented as second-order differential equations, this force should be
expressed in a functional form such as that given on the right-side of Eq.~%
\eqref{eq4.1a}.

An equation of particular importance for the modeling of a broad range of
nonlinear oscillatory phenomena is the Duffing equation. \cite{6} It
provides a useful and realistic model for pendulum-type problems, and for a
variety of issues involving nonlinear vibrations/oscillations of beams,
cables, electrical circuits, and plates; see Kovacic and Brennan, Chapter~2.
Another recent application of the Duffing equation is to the analysis and
understanding of the mechanical and electro-mechanical properties of
graphene. \cite{10,11} Single and multi-sheets of this material are being
studied for a wide range of possible measuring devices which may be used in
new technological applications.

In standard form the Duffing equation can be written as 
\begin{equation}  \label{eq4.2}
m\ddot x + \epsilon f(\dot x)+\omega^2x+\beta_1x^2+\beta_2x^3=0
\end{equation}
for the the case where there is no external driving fore; \cite{2,3,6} $%
(m,\epsilon,\omega^2)$ are non-negative, while $(\beta_1,\beta_2)$ may be of
either sign or zero. However, there are also important systems where the
sign of the $\omega^2x$ term is taken to be negative. \cite{6} To date, the
major emphasis has been on the study of Eq.~\eqref{eq4.2} where $f(v)$ is
that given by Eq.~\eqref{eq2.2}, with $\alpha=1$ or 2; see Ref.~%
\cite{6}.

For forced oscillations, a large research literature exists on the case
where a forcing term $F\cos (\Omega t)$ is placed on the right-side of Eq.~%
\eqref{eq4.2}, i.e., 
\begin{equation}
m\ddot{x}+\epsilon f(\dot{x})+\omega ^{2}x+\beta _{1}x^{2}+\beta
_{2}x^{3}=F\cos (\Omega t),  \label{eq4.3a}
\end{equation}%
where $\Omega $ is the external driving (angular) frequency. Most of these
studies are based on using $\alpha =1$ for the damping force. \cite{2,3,4,6}
For other positive, integer values, a technique called equivalent viscous
damping may be used to effectively linearize the nonlinear damping force
(see Ref.\cite{6}; section 6.2.3 and Ref.\cite{12}). In
general, the major outcome of these calculations is that different values of 
$\alpha $ produce differences in the properties of the forced oscillatory
states (see Ref. \cite{6}, section 6.8). However, the overall
qualitative features for $\alpha $ integer, but greater than one, is similar
to the case where $\alpha =1$.

Based on the results of our study, it might be useful to consider values of $%
\alpha $ belonging to the interval $(0,1)$ for the modeling of physical
systems having damping. Note that if $\alpha $ is selected to be 
\begin{equation}
\alpha =%
\begin{cases}
\frac{2n+1}{2m+1}, \\ 
(n,m)\mbox{ non-negative integer}, \\ 
n<m,%
\end{cases}
\label{eq4.3}
\end{equation}%
then the damping function 
\begin{equation}
f(v)=\lambda v^{\frac{2n+1}{2m+1}},  \label{eq4.4}
\end{equation}%
is automatically an odd function \cite{13} and the $\mathop{\rm sgn}(v)$
function, see Eq.~\eqref{eq2.2}, is not required in the specification of $%
f(v)$. Elastic force functions corresponding to the case, $n=0$ and $m=1$,
has been extensively investigated for the case of undamped oscillators \cite%
{13}. The simplest type of such an oscillator is \cite{13} 
\begin{equation}
\ddot{x}+x^{1/3}=0.  \label{eq4.5}
\end{equation}%
An extension of this type of nonlinear term to the damping force produces
the following version of the Duffing equation 
\begin{equation}
m\ddot{x}+\epsilon \dot{x}^{1/3}+\omega ^{2}x+\beta _{1}x^{2}+\beta
_{2}x^{3}=F\cos (\Omega t).  \label{eq4.6}
\end{equation}%
It would be of considerable theoretical and practical value to investigate
the unforced version of Eq.~\eqref{eq4.6}, with $\beta _{1}\geq 0$ and $%
\beta _{2}>0$, and determine whether the damped oscillations decay to zero
in a finite time. If so, this would provide justification to not use damping
forces depending solely on integer powers of $\alpha $. Further, this result
would then require a detailed analysis of the physical mechanisms relating
to various damping forces which might give values of $\alpha $ such that $%
0<\alpha <1$. In particular, a very important result would be to determine
whether $\alpha $ must take on the fractional values indicated in Eq.~%
\eqref{eq4.3}, or just be a real number.

It should be noted that previous work by Mickens et al. \cite{14} on a
cube-root damped harmonic oscillator 
\begin{equation}
\ddot{x}+x=-2\epsilon \dot{x}^{1/3},
\end{equation}%
showed that the oscillations decreased to zero in a finite time. This
implies that only a finite number of oscillations take place before the
equilibrium state, $x=0$, is reached. Exactly, the same conclusion is
reached for the singular case, i.e., $\alpha =0$, which represents dry
friction or Coulomb damping, \cite{2,3,4} and for the other cases where the $%
\frac{1}{3}$ in Eq. (30), is replaced by $\alpha $, such that $0<\alpha <1.$

\section{\protect\bigskip Conclusions}

\label{sec5} Within the context of a model system, in which the only force
acting on a particle is a damping force, it has been shown that a class of
dissipative forces exist such that the dynamic behavior ends in a finite
time. An important aspect of this result is that it agrees with what is
observed for the actual behavior of physical systems, i.e., the oscillations
or motions stop in a finite time. We believe the use of such damping forces
in the construction of mathematical models of physical systems may provide
more realistic models than are currently given by using a linear \ or
quadratic (in velocity) damping term.

Finally, it should be indicated that in the \ modeling of a single particle
system, in one space dimensions, a very general damping and/or dissipative
force takes the form%
\begin{equation}
f(x,v)=\left[ sgn(v)\right] \left[ a_{1}\left\vert v\right\vert ^{\alpha
}+a_{2}\left\vert v\right\vert ^{2}\right] +\left[
b_{1}+b_{2}x^{2}+b_{3}v^{2}\right] v,  \label{eq5.1}
\end{equation}%
where the parameters $(a_{1},a_{2},b_{1},b_{2},b_{3})$ are all non-negative,
and $0<\alpha <1.$ This functional form includes all of the standard
proposed power-law laws, including the fractional power force.\cite{1, 3, 5,
6} By applying the method of"dominant balance" \cite{15}, it follows that \
for small v%
\begin{equation}
f(x,v)\simeq a_{1}\left[ sgn(v)\right] \left\vert v\right\vert ^{\alpha },
\label{eq5.2}
\end{equation}%
and, consequently, finite time dynamics is expected to result.

\section*{acknowledgments}
The work reported in this document was supported in part by Faculty Development Funds from the Office of the Dean, School of Arts and Sciences, Clark Atlanta University. 


\begin{thebibliography}{99}
\bibitem{1} V. Lazan, Damping of Materials and Members in Structural
Mechanics, (Pergamon, New York, 1968). See section~3.1.

\bibitem{2} A. H. Nayfeh, D. T. Mook, Nonlinear Oscillations,
(Wiley-Interscience, New York, 1979).

\bibitem{3} G. Schmidt, A. Tondl, Non-Linear Vibrations, (Cambridge
University Press, New York, 1986). See pps.~25 and 52.

\bibitem{4} R. E. Mickens, Nonlinear Oscillations, (Cambridge University
Press, New York, 1981). See section~1.2.7.

\bibitem{5} N. de Mestre, The Mathematics of Projectiles in Sport,
(Cambridge University Press, New York, 1990). See sections~3.1, 3.4, and 3.6.

\bibitem{6} A. K. Mallik in: I. Kovacic, M. J. Brennan (Eds.), The Duffing
Equation, (Wiley, Chichester, 2011). See chapters~3 and 6.

\bibitem{7} S. L. Ross, Differential Equations, second ed., Xerox,
(Lexington, MA, 1974).

\bibitem{8} A. Jeffrey, Handbook of Mathematical Formulas and Integrals,
second ed., (Academic, New York, 2000).

\bibitem{9} G. Strang, Calculus, (Wellesley-Cambridge Press, Wellesley, MA,
1991). See section~7.5.

\bibitem{10} A. K. Geim, K. S. Novoselov, "The Rise of Graphene" Nature
Materials 6 (2007) 183--191.

\bibitem{11} L. C. Shao, M. Palaniapan, W. W. Tan, L. Khine, "Nonlinearity
in Micromechanical free-free beam resonators: Modeling and Experimental
Verification" J. Micromech. Microeng. 18 (2008)
doi:10.1088/0960-1317/18/2/025017

\bibitem{12} J. E. Ruzicka, T. F. Derby, Influence of Damping in Vibration
Isolation, Naval Research Laboratory, Washington, DC 1971.

\bibitem{13} R. E. Mickens, Truly Nonlinear Oscillations, (World Scientific,
Singapore, 2010).

\bibitem{14} \ R. E. Mickens, K. O. Oyedeji, S. A. Rucker, \textquotedblleft
Analysis of the simple harmonic oscillator with fractional
damping\textquotedblright\ J. Sound Vib. 268 (2003) 839--842.

\bibitem{15} C. M. Bender, S. A. Orszag, Advanced Mathematical Methods for
Scientists and Engineers, (McGraw-Hills, New York, 1978). See pps. 83-88.
\end{thebibliography}
\end{document}